\documentclass[12pt,aps,prl]{revtex4}
\usepackage{epsfig}
\usepackage{amsmath}

\normalsize

\def\bk{{\bf k}}

\def\bq{{\bf q}}

\def\bv{{\bf v}}

\def\b0{{\bf 0}}

\def\Im{{\rm Im}}

\def\eps{\epsilon}
\def\gam{\gamma}
\def\Gam{\Gamma}
\def\lam{\lambda}
\def\Lam{\Lambda}
\def\om{\omega}
\def\Om{\Omega}
\def\sg{\sigma}
\def\Sg{\Sigma}

\begin{document}


\title{Electrical resistivity near Pomeranchuk instability \\ 
       in two dimensions}

\author{Luca Dell'Anna$^{1,2}$ and Walter Metzner$^1$}

\affiliation{$^1$Max-Planck-Institute for Solid State Research,
 D-70569 Stuttgart, Germany \\
 $^2$Institut f\"ur Theoretische Physik, Heinrich-Heine Universit\"at,
 D-40225 D\"usseldorf, Germany}

\date{\small\today}


\begin{abstract}
We analyze the DC charge transport in the quantum critical regime 
near a d-wave Pomeranchuk instability in two dimensions.
The transport decay rate is linear in temperature everywhere on the 
Fermi surface except at cold spots on the Brillouin zone diagonal.
For pure systems, this leads to a DC resistivity proportional
to $T^{3/2}$ in the low-temperature limit. 
In the presence of impurities the residual impurity resistance 
at $T=0$ is approached linearly at low temperatures.

\noindent
\mbox{PACS: 71.10.Hf, 72.10.Di} \\
\end{abstract}

\maketitle

Pomeranchuk instabilities \cite{Pom} leading to symmetry-breaking
deformations of the Fermi surface in interacting electron systems
have attracted much interest in the last few years. 
Interactions favoring a Pomeranchuk instability with d-wave symmetry 
have been found in the two most intensively studied single-band models 
for cuprate superconductors, that is, the two-dimensional t-J 
\cite{YK1} and Hubbard \cite{HM,GKW} model.
These models thus exhibit enhanced ''nematic'' correlations, 
as usually discussed in the context of fluctuating stripe order 
\cite{KFE}.
Signatures for incipient nematic order with d-wave symmetry have 
been observed in various cuprate materials \cite{KBX}. In particular, 
nematic correlations close to a d-wave Pomeranchuk instability
provide a natural explanation for the relatively strong in-plane 
anisotropy observed in the magnetic excitation spectrum of 
$\rm Y Ba_2 Cu_3 O_y$ \cite{HPX,YM}.
A spin dependent Pomeranchuk instability was recently invoked 
to explain a new phase observed in ultrapure crystals of the 
layered ruthenate metal $\rm Sr_3 Ru_2 O_7$ \cite{GGX},
and also to account for a puzzling phase transition in
$\rm U Ru_2 Si_2$ \cite{VZ}.

Critical fluctuations near a Pomeranchuk instability provide an
interesting route to non-Fermi liquid behavior in two dimensions
\cite{OKF,MRA}.
The properties of single-particle excitations near a 
quantum critical point associated with a Pomeranchuk instability 
have been studied already in considerable detail \cite{DM,RPC}.
For a d-wave Pomeranchuk instability in an electron system on a 
square lattice the singular part of the electronic self-energy
is proportional to $d_{\bk}^2$, where $d_{\bk}$ is a form factor
with d-wave symmetry \cite{MRA}.
At the quantum critical point, the real and imaginary parts of 
the self-energy scale as $|\om|^{2/3}$ with energy \cite{OKF,MRA}. 
This leads to a complete destruction of quasi-particles near the
Fermi surface except for the ''cold spots'' on the Brillouin zone 
diagonal, where the form factor $d_{\bk}$ vanishes.
In the quantum critical regime at $T>0$ the self-energy consists 
of a ''classical'' and a ''quantum'' part with very different 
dependences on $T$ and $\om$. 
The classical part, which is due to classical fluctuations,
dominates at $\om = 0$ and yields a contribution proportional 
to $\sqrt{T/\log T}$ to the imaginary part of the self-energy 
on the Fermi surface \cite{DM}.

In this letter we compute the temperature dependence of the DC
resistivity in the quantum critical regime near a Pomeranchuk
instability in two dimensions.
We obtain a momentum dependent transport decay rate 
$\gam^{\rm tr}_{\bk}(T)$ which is {\em linear} in temperature 
for all momenta on the Fermi surface except at the cold spots 
on the Brillouin zone diagonal.
Adding a conventional $T^2$-term to $\gam^{\rm tr}_{\bk}(T)$ 
we obtain an overall resistivity $\rho(T)$ proportional to 
$T^{3/2}$ at low temperatures. 
In the presence of impurities, the residual resistivity at
zero temperature is approached linearly.

Our calculations are based on a phenomenological lattice model
\cite{MRA},
\begin{equation}
 H = H_0 +
 \frac{1}{2V} \sum_{\bk,\bk',\bq} f_{\bk\bk'}(\bq) \,
 n_{\bk}(\bq) \, n_{\bk'}(-\bq) \; ,
\end{equation}
where $H_0$ is a kinetic energy, $n_{\bk}(\bq) = 
 \sum_{\sg} c^{\dag}_{\bk-\bq/2,\sg} c_{\bk+\bq/2,\sg}$, 
and $V$ is the volume of the system.
Since the Pomeranchuk instability is driven by interactions with
small momentum transfers (forward scattering), we choose 
a coupling function $f_{\bk\bk'}(\bq)$ which contributes only 
for relatively small momenta $\bq$. This excludes other 
instabilities such as superconductivity or density waves. 
We consider an interaction of the form \cite{DM,YOM}
\begin{equation}
 f_{\bk\bk'}(\bq) = u(\bq) + g(\bq) \, d_{\bk} \, d_{\bk'} 
\end{equation}
with $u(\bq) \geq 0$ and $g(\bq) < 0$, and a form factor $d_{\bk}$ 
with $d_{x^2-y^2}$ symmetry, such as $d_{\bk} = \cos k_x - \cos k_y$.
The coupling functions $u(\bq)$ and $g(\bq)$ vanish if 
$|\bq|$ exceeds a certain small momentum cutoff $q_c$.
This ansatz mimics the effective interaction in the forward 
scattering channel as obtained from renormalization group 
calculations \cite{HM} for the two-dimensional Hubbard model.
The uniform term originates directly from the repulsion between
electrons and suppresses the electronic compressibility of the 
system. The d-wave term drives the Pomeranchuk instability.

The Pomeranchuk instability can actually be preempted by a 
first order transition at low temperatures, where the Fermi 
surface symmetry changes abruptly before the fluctuations 
become truely critical \cite{KKC}. 
However, for reasonable choices of hopping and interaction 
parameters the system is nevertheless characterized by 
strong fluctuations on the symmetric side of the transition 
\cite{YOM}.
The first order character of the transition is suppressed 
by the uniform repulsion $u$ in (2), and for a favorable but 
not unphysical choice of model parameters a genuine quantum 
critical point can be realized \cite{YOM}.

Near the Pomeranchuk instability, the electrons interact 
via a singular effective interaction of the form \cite{MRA,DM}
\begin{equation}
 \Gam_{\bk\bk'}(\bq,\nu) = \frac{g \, d_{\bk} d_{\bk'}}
 {(\xi_0/\xi)^2 + \xi_0^2|\bq|^2 - i[\nu/(c|\bq|)]} \; ,
\end{equation}
where $\bq$ and $\nu$ is the momentum and energy transfer,
respectively. The parameters $g = g(\b0)$, $\xi_0$ and 
$c$ can be treated as constants, whereas the correlation 
length $\xi$ depends sensitively on control parameters and 
temperature.
In the quantum critical regime $\xi(T)$ is proportional to
$(T|\log T|)^{-1/2}$. 

The electron self-energy $\Sg(\bk,\om)$ has been computed 
previously \cite{MRA,DM,RPC} in random phase approximation 
(RPA) with the effective interaction (3).
In the quantum critical regime one finds \cite{DM}
\begin{equation}
 \Im\Sg(\bk_F,0) = 
 \frac{g \, d_{\bk_F}^2}{4 v_{\bk_F} \xi_0^2} \, T \xi(T)
\end{equation}
for momenta $\bk_F$ on the Fermi surface 
($v_{\bk_F}$ is the Fermi velocity).
The corresponding approximation for the electrical resistivity 
involves the RPA self-energy and current vertex corrections 
due to particle-hole ladder diagrams, in close analogy to the 
Born approximation for impurity scattering \cite{Ric}.
We assume that the Pomeranchuk fluctuations thermalize
sufficiently rapidly such that the effective interaction
(3) is not modified by the electric current.
This relaxation to equilibrium is not described by the model
(1) and has to be provided by additional terms such as
umklapp, impurity or phonon scattering \cite{CGDE,LRVW}.

The DC conductivity (inverse resistivity) can be obtained from
the retarded current-current correlation function $\Pi$ as
\begin{equation}
 \sg_{jj'} = - \lim_{\om \to 0} \lim_{\bq \to \b0} 
 \frac{e^2}{\om} \, \Im\Pi_{jj'}(\bq,\om)
\end{equation}
For cubic symmetry the conductivity tensor is diagonal.
The current-current correlator can be expressed (exactly) in 
terms of single-particle Green functions $G$ and current 
vertices.
Performing an analytic continuation from Matsubara to real 
frequencies and taking the DC limit, one obtains
\begin{equation}
 \sg_{jj'} = 
 - \frac{e^2}{\pi} \int d\om \, f'(\om) \int \frac{d^2k}{(2\pi)^2} \,
 \Lam^0_j(\bk) \, |G(\bk,\om)|^2 \, \Lam_{j'}(\bk,\om) \; ,
\end{equation}
where $f(\om)$ is the Fermi function,
${\bf\Lam}^0(\bk) = \bv_{\bk} = \nabla\eps_{\bk}$ the bare current 
vertex, and ${\bf\Lam}(\bk,\om)$ is the interacting current vertex 
in the mixed advanced-retarded DC limit, that is,
${\bf\Lam}(\bk,\om) = {\bf\Lam}(\bk,\om+i0^+;\bk,\om-i0^+)$.
The product $|G|^2$ can be expressed in terms of the single-particle
spectral function $A(\bk,\om) = -\frac{1}{\pi} \Im G(\bk,\om)$ and
the (retarded) self-energy as
\begin{equation}
 |G(\bk,\om)|^2 = 
 - \frac{\pi A(\bk,\om)}{\Im\Sg(\bk,\om)} \; .
\end{equation}
At low temperatures the derivative of the Fermi function $f'(\om)$ 
has a sharp peak of width $T$ at $\om = 0$. Since all other factors
under the integral  in Eq.\ (6) have a broader $\om$-dependence, one
can replace $f'(\om)$ by $-\delta(\om)$, such that the conductivity
can be written as
\begin{equation}
 \sg_{jj'} = 
 - e^2 \int \frac{d^2k}{(2\pi)^2} \,
 \Lam^0_j(\bk) \, \frac{A(\bk,0)}{\Im\Sg(\bk,0)}
 \Lam_{j'}(\bk,0) \; .
\end{equation}

The interacting current vertex includes all particle-hole ladder
vertex corrections. It is thus obtained from a linear integral
equation, which can be written as
\begin{eqnarray}
 {\bf\Lam}(\bk,\om) &=& 
 {\bf\Lam}^0(\bk) + \int d\eps \int \frac{d^2q}{(2\pi)^2} \, 
 [b(\eps) + f(\om+\eps)] \nonumber \\
 &\times& \Im\Gam_{\bk\bk}(\bq,\eps) \,
 \frac{A(\bk+\bq,\om+\eps)}{\Im\Sg(\bk+\bq,\om+\eps)} \,
 {\bf\Lam}(\bk+\bq,\om+\eps)
\end{eqnarray}
after analytic continuation to the real frequency axis. 
Here $b(\eps)$ is the Bose function.

For $T \to 0$ the correlation length $\xi(T)$ diverges.
Repeating the arguments used for the calculation of $\Sg(\bk,\om)$ in 
Ref.~\onlinecite{DM}, one finds that the integration variable $\eps$ 
in Eq.~(9) scales as $\xi^{-3}$ and can therefore be set to zero in 
the arguments of $A$, $\Sg$, and ${\bf\Lam}$ on the right hand side 
of Eq.~(9). 
Expanding the Bose function as $b(\eps) \sim T/\eps$, one can carry
out the $\eps$-integration explicitly,
$\int d\eps \, \eps^{-1} \, \Im\Gam_{\bk\bk}(\bq,\eps) \, =
 \pi \Gam_{\bk\bk}(\bq,0)$, yielding a closed equation for the
{\em static}\/ current vertex ${\bf\Lam}(\bk) = {\bf\Lam}(\bk,0)$
\begin{equation}
 {\bf\Lam}(\bk) = 
 {\bf\Lam}^0(\bk) + T \int \frac{d^2q}{(2\pi)^2} \Gam_{\bk\bk}(\bq,0) 
 \, \frac{\pi A(\bk+\bq,0)}{\Im\Sg(\bk+\bq,0)} \, {\bf\Lam}(\bk+\bq)
 \; .
\end{equation}
The same result could have been obtained by considering only the 
{\em classical}\/ fluctuations, that is, by including only the term
$\Gam_{\bk\bk}(\bq,i\eps_n)$ with Matsubara frequency $\eps_n = 0$
in the Matsubara sums for the current vertex corrections.
At this point the equations for $\sg_{jj'}$ and ${\bf\Lam}(\bk)$ are 
formally identical to those obtained from the Born approximation in 
{\em disordered}\/ electron systems with a $\bk$-dependent long-ranged 
disorder correlator given by $\Gam_{\bk\bk}(\bq,0)$.

Inserting the ansatz ${\bf\Lam}(\bk) = \lam(\bk) \bv_{\bk}$ into the
equation for the current vertex, one obtains the following equation 
for the function $\lam(\bk)$:
\begin{equation}
 {\bf\lam}(\bk) = 
 1 + T \int \frac{d^2q}{(2\pi)^2} \Gam_{\bk\bk}(\bq,0) 
 \, \frac{\pi A(\bk+\bq,0)}{\Im\Sg(\bk+\bq,0)} \, 
 \frac{\bv_{\bk} \cdot \bv_{\bk+\bq}}{v_{\bk}^2} \, {\bf\lam}(\bk+\bq)
 \; .
\end{equation}
Since the conductivity is dominated by momenta near the Fermi surface,
we now focus on the case $\bk=\bk_F$.
For large $\xi$ the above integral is dominated by small momentum 
transfers $\bq$ of order $\xi^{-1}$, due to the effective interaction
$\Gam_{\bk\bk}(\bq,0)$.
The spectral function is peaked for momenta on the Fermi surface,
with a width determined by $\Im\Sg(\bk_F,0)$, which is proportional
to $T\xi(T)$.
The self-energy $\Sg(\bk,0)$ varies on a momentum scale of order 
$\xi^{-1}$ for momentum shifts perpendicular to the Fermi surface 
\cite{DM}. 
The same can be expected for $\lam(\bk)$, since the current vertex
correction can be related to the shift of the self-energy in the
presence of a field coupled to the current operator. 
Since $T\xi^2(T) \propto 1/\log T$ in the quantum critical regime,
and since the tangential $\bq$-dependence of $\Im\Sg(\bk_F+\bq)$ 
and $\lam(\bk_F+\bq)$ is negligible on the scale $\xi^{-1}$,
we may neglect the $\bq$-dependence of $\Im\Sg(\bk_F+\bq)$ and 
$\lam(\bk_F+\bq)$ in (11) altogether, which can then be solved 
explicitly, yielding
\begin{equation}
 {\bf\lam}(\bk_F) = \left[ 
 1 - \frac{\pi T}{\Im\Sg(\bk_F,0)} 
 \int \frac{d^2q}{(2\pi)^2} \, \Gam_{\bk_F\bk_F}(\bq,0) 
 \, A(\bk_F+\bq,0) \, 
 \frac{\bv_{\bk_F} \cdot \bv_{\bk_F+\bq}}{v_{\bk_F}^2} \right]^{-1}
 \; .
\end{equation}
Using 
\begin{equation}
\Im\Sg(\bk_F,0) = \pi T \int \frac{d^2q}{(2\pi)^2} \,
 \Gam_{\bk_F\bk_F}(\bq,0) \, A(\bk_F+\bq,0) \; ,
\end{equation}
which is true within self-consistent RPA restricted to classical 
fluctuations \cite{DM}, one can write $\lam_{\bk_F}$ as
\begin{equation}
 \lam(\bk_F) = \gam_{\bk_F}/\gam^{\rm tr}_{\bk_F} \; ,
\end{equation}
where $\gam_{\bk_F} = - \Im\Sg({\bk_F,0})$ is the single-particle
decay rate while
\begin{equation}
 \gam^{\rm tr}_{\bk_F} = - \pi T 
 \int \frac{d^2q}{(2\pi)^2} \, \Gam_{\bk_F\bk_F}(\bq,0) 
 \, A(\bk_F+\bq,0) \, \left( 1 -
 \frac{\bv_{\bk_F} \cdot \bv_{\bk_F+\bq}}{v_{\bk_F}^2} \right)
 \; .
\end{equation}
is the scattering rate relevant for transport.

The momentum integral in the expression (8) for the conductivity 
is peaked at the Fermi surface. 
For $T \to 0$ with $\xi(T) \propto (T \log T)^{-1/2}$ one can
replace $A(\bk,0)$ under the integral by 
$\delta(\eps_{\bk} - \mu)$, neglecting possible corrections 
of order $1/\log T$, such that the conductivity can be written as
a Fermi surface integral. 
Inserting Eq.~(14) for $\lam(\bk_F)$, one obtains
\begin{equation}
 \sg = \frac{e^2}{8\pi^2} \int d\Om_{\bk_F} \, 
 \frac{v_{\bk_F}}{\gam^{\rm tr}_{\bk_F}}
\end{equation}
for the diagonal part $\sg = \sg_{jj}$ of the conductivity tensor.

To compute $\gam^{\rm tr}_{\bk_F}$, we parametrize the (small) momentum
transfer $\bq$ in Eq.~(15) by radial and tangential components,
$q_r$ and $q_t$, respectively. 
For $T \to 0$, we may again approximate $A(\bk_F+\bq,0)$ by a 
$\delta$-function, $\delta(\eps_{\bk_F+\bq} - \mu)$.
The dispersion relation can be expanded as
$\eps_{\bk_F+\bq} - \mu = v_{\bk_F} q_r + q_t^2/(2m^t_{\bk_F})$,
where $(m^t_{\bk_F})^{-1} = 
\left. \partial_{k_t}^2 \eps_{\bk} \right|_{\bk_F} \,$.
Since $\bk_F+\bq$ is confined to the Fermi surface, the momentum 
transfers $\bq$ are predominantly tangential to the Fermi surface
in $\bk_F$, such that the term of order $q_t^2$ cannot be neglected
compared to the term linear in $q_r$. 
After expanding also the kinematic factor 
$1 - \frac{\bv_{\bk_F} \cdot \bv_{\bk_F+\bq}}{v_{\bk_F^2}}$ to linear 
order in $q_r$ and quadratic order in $q_t$, the momentum integral 
in Eq.~(15) can be performed analytically.
For $\xi^{-1}(T) \ll m^t_{\bk_F} v_{\bk_F}$, one obtains
\begin{equation}
 \gam^{\rm tr}_{\bk_F} = 
 \frac{|g|}{\pi \xi_0^2} \, m^t_{\bk_F}
 \arctan\Big(\frac{q_c}{2m^t_{\bk_F} v_{\bk_F}} \Big) \,
 K_{\bk_F} d_{\bk_F}^2 \, T \; ,
\end{equation}
where $q_c$ is the momentum cutoff and
\begin{equation}
 K_{\bk_F} = \frac{1}{2v_{\bk}^2} \left. \left(
 \frac{\bv_{\bk} \cdot \partial_{k_r} \bv_{\bk}}
 {v_{\bk} \, m^t_{\bk}} -
 \bv_{\bk} \cdot \partial^2_{k_t} \bv_{\bk}
 \right) \right|_{\bk=\bk_F} \; .
\end{equation}
The function $K_{\bk_F}$ has units of inverse momentum 
squared. For a quadratic dispersion relation,
$\eps_{\bk} = k^2/(2m)$, one has $m^t_{\bk} = m$ and
$K_{\bk_F} = 1/(2k_F^2)$.
The scattering rate $\gam^{\rm tr}_{\bk_F}$ is thus {\em linear}
in $T$ at low temperatures. 
Note that the correlation length $\xi(T)$ does not appear in the 
asymptotic low temperature behavior of $\gam^{\rm tr}_{\bk_F}$.
This behavior in the quantum critical regime can be contrasted
with the behavior in the Fermi liquid regime close to the quantum 
critical point. For the latter case $\gam^{\rm tr}_{\bk_F}$ is 
proportional to $\xi^2 T^2 \log T$, 
where $\xi = \xi(T \to 0)$ \cite{LRVW}. 

Due to the d-wave form factor the prefactor of the $T$-linear 
behavior of $\gam^{\rm tr}_{\bk_F}$ varies strongly along the 
Fermi surface and vanishes on the Brillouin zone diagonal. 
This is reminiscent of the {\em cold spot}\/ scenario of transport 
in cuprates \cite{IM}.
Inserting $\gam^{\rm tr}_{\bk_F}$ from (17) in Eq.~(16) for the
conductivity one obtains a divergent Fermi surface integral,
due to the zeros of $\gam^{\rm tr}_{\bk_F}$ at the cold spots
$\bk_F^c$. 
In the absence of any other scattering mechanism, the 
conductivity would thus be infinite.
However, other (than d-wave forward scattering) residual 
interactions will lead at least to the conventional Fermi
liquid decay rate of order $T^2$ all over the Fermi surface,
including the cold spots.
Including a Fermi liquid term of order $T^2$, the scattering
rate has the form
$\gam^{\rm tr}_{\bk_F}(T) = 
 a_{\bk_F} T^2 + b_{\bk_F} d_{\bk_F}^2 T$, 
where the coefficients $a_{\bk_F}$ and $b_{\bk_F}$ are 
finite for all $\bk_F$.
Inserting this ansatz into Eq.~(16), one finds a resistivity
\begin{equation}
 \rho(T) = \frac{2\pi}{e^2} \, 
 \frac{\sqrt{a_{\bk_F^c} b_{\bk_F^c}}}{v_{\bk_F^c}} \, 
 T^{3/2}
\end{equation}
for low $T$. 
Taking the well-known logarithmic correction to the $T^2$-behavior 
of the scattering rate in two-dimensional Fermi liquids into 
account, one obtains 
$\rho(T) \propto T^{3/2} \, |\log T|^{1/2} \,$.

In the presence of impurities, the scattering rate has the form
$\gam^{\rm tr}_{\bk_F}(T) = 
 \gam^{\rm imp}_{\bk_F} + b_{\bk_F} d_{\bk_F}^2 T$, for 
temperatures low enough that the Fermi liquid term of order
$T^2$ can be neglected compared to the impurity term.
For $T \to 0$ one then obtains a finite residual resistivity
determined exclusively by impurity scattering.
For low finite temperatures the resistivity increases 
linearly with $T$ as long as 
$T \ll \gam^{\rm imp}_{\bk_F}/b_{\bk_F}$.
For $T \gg \gam^{\rm imp}_{\bk_F}/b_{\bk_F}$ one obtains
$\rho(T) \propto T^{1/2}$, with a prefactor proportional to
$(\gam^{\rm imp}_{\bk_F^c} \, b_{\bk_F^c})^{1/2}/v_{\bk_F^c} \,$, 
provided that impurity scattering still dominates over the 
conventional Fermi liquid contribution to 
$\gam^{\rm tr}_{\bk_F}(T)$.

In summary, we have analyzed the DC charge transport in the 
quantum critical regime near a d-wave Pomeranchuk instability 
in two dimensions.
It turned out that the relaxation of the electric current is
dominated by classical fluctuations.
The transport decay rate $\gam^{\rm tr}_{\bk_F}(T)$ is linear
in temperature everywhere on the Fermi surface except at cold 
spots on the Brillouin zone diagonal.
For pure systems, this leads to a DC resistivity proportional
to $T^{3/2}$ in the low-temperature limit. 
It is tempting to associate this result with the unusual
$T^{3/2}$-law observed for the resistivity in overdoped 
$\rm La_{2-x} Sr_x Cu O_4$ \cite{TBK}. 
In the presence of impurities the residual impurity resistance 
at $T=0$ is approached linearly at low temperatures.

\vskip 1cm

\noindent
{\bf Acknowledgements:} We are grateful to C. Castellani, 
 A. Chubukov, C. Di Castro, A.~Katanin, A. Rosch, P. W\"olfle,
 and H. Yamase for valuable discussions.


\vfill\eject

\end{document}